\documentclass[aps,prl,twocolumn,groupedaddress]{revtex4-2}

\usepackage{graphicx}
\usepackage{amsmath,amssymb}
\usepackage{hyperref}

\begin{document}

\title{A $^{3}$He-$^{21}$Ne Ramsey Comagnetometer with sub-nHz frequency resolution}

\author{Shaobo Zhang}
\author{Jingyao Wang}
\author{George Sun}
\author{Johannes J. van de Wetering}
\author{Michael V. Romalis}
\affiliation{Department of Physics, Princeton University, Princeton, New Jersey 08544, USA}
\date{\today}

\begin{abstract}
Nuclear spin comagnetometers offer exceptional precision in measurements of spin energy levels and exhibit long-term stability, making them powerful tools for probing spin-dependent physics beyond the Standard Model as well as for inertial rotation sensing. We describe a new $^{3}$He-$^{21}$Ne Ramsey comagnetometer operating with an in-situ $^{87}$Rb magnetometer for initialization and sensing of nuclear spins. During free precession of nuclear spins we turn off all lasers and introduce a microwave field to suppress back-action from Rb atoms. We demonstrate that scalar and dipolar interactions between nuclear spins can be eliminated via control of the polarized sample geometry. These improvements result in a bias-free measurement with a frequency sensitivity of 0.6 nHz after 6 hours of integration.

\end{abstract}

\maketitle

Precision measurements of spin-dependent phenomena have emerged as a powerful probe of physics beyond the Standard Model~\cite{Safronova_2018,Terrano_2022,Cong_2025}.  Nuclear spin precession measurements provide one of the most sensitive methods for detection of absolute energy shifts between quantum states, reaching energy resolutions below $10^{-25}$ eV~\cite{Graner_2016,Vasilakis_2009}. By simultaneously monitoring the spin precession frequencies of two nuclear species in a shared volume and suppressing common-mode magnetic noise~\cite{Lamoreaux_86}, comagnetometers can set limits on violations of fundamental symmetries, such as CP \cite{EDM_chupp,EDM_Heil}, CPT \cite{Allmendinger_2014, CPT_Brown}, and Lorentz invariance \cite{Smiciklas_2011}. They are also used to search for new long-range forces~\cite{Cong_2025,Hunter_2013,Kimball_2023,Zhang_2023} and for axion dark matter \cite{Lee_axion,Budker_axion}.

In addition to their role in fundamental physics, comagnetometers can be used for inertial rotation sensing~\cite{Woodman1987,Kornack_2005,Walker2016,Wei_2023} by utilizing nuclear spins that are relatively isolated from the environment as a quantum gyroscope. They  achieve sensitivity comparable to that of atom interferometer gyroscopes \cite{Geiger2020}, while benefiting from a simpler construction and an intrinsically stable scale factor.

\begin{figure*}[t]
  \centering
  \includegraphics[width=0.9\textwidth]{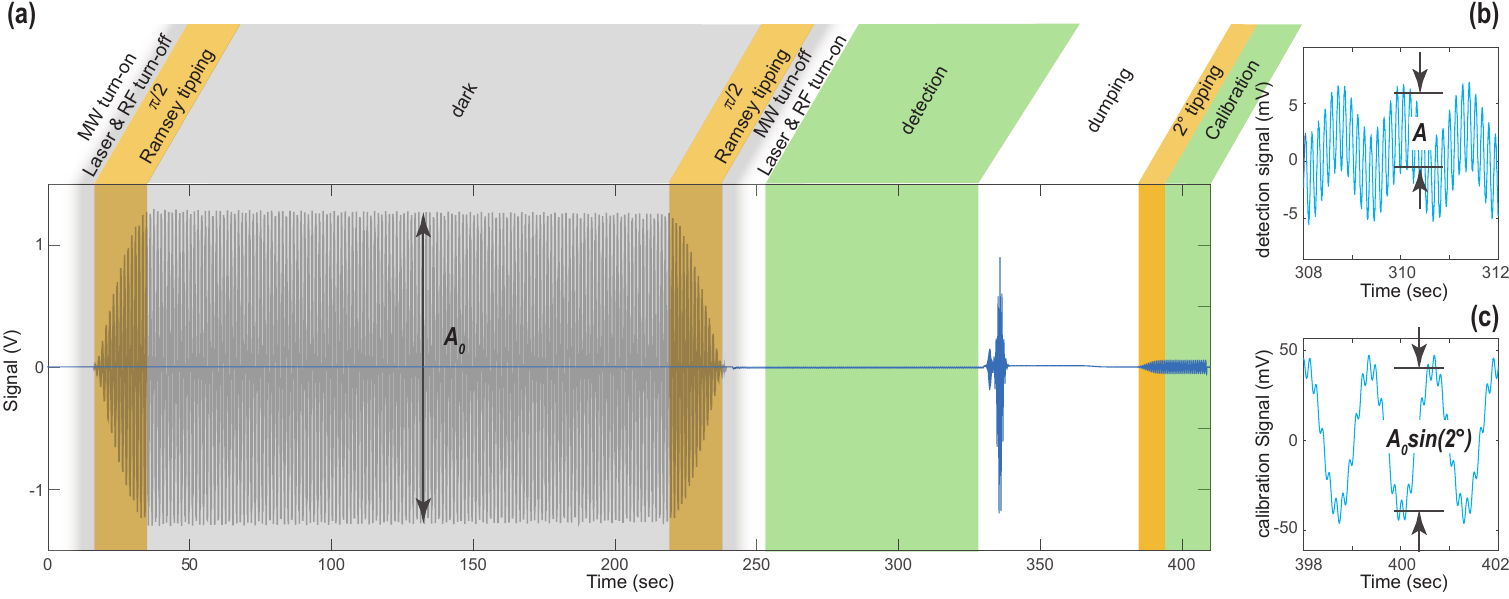}
  \caption{(a) Schematic of the in-the-dark operation of the nuclear spin comagnetometer. The gray solid lines illustrate the evolution of the two nuclear spins under the Ramsey scheme, while blue lines indicate recorded signals. The detection time is approximately equal to half of the free precession time to yield the optimal Cramér–Rao lower bound.
(b) and (c) Nuclear spin precession signals during the detection stage and after a $2^\circ$ calibration pulse. The amplitudes $A_0$ and $A$ are obtained separately for each nuclear spin, with only the \textsuperscript{21}Ne amplitude indicated in the figure.}
  \label{fig:fullwidth}
\end{figure*}

The most sensitive comagnetometers to date operate in the \textit{self-compensating} regime, where nuclear spins are coupled to alkali-metal electron spins via spin-exchange interactions~\cite{Kornack_2002,Vasilakis_2009,Wei_2023}. By tuning the external magnetic field to cancel the effective field from noble gas magnetization, these systems automatically suppress magnetic noise and isolate non-magnetic signals. This configuration offers  good short-term sensitivity but suffers from long-term drift due to laser intensity and pointing fluctuations~\cite{Wang_2025}. In addition, self-compensating comagnetometers are fundamentally limited by the sensitivity of the underlying alkali-metal magnetometer. 

In contrast, \textit{clock-comparison} comagnetometers utilize two freely precessing nuclear spin species with the signal detected optically~\cite{Lamoreaux_86}, with a pick-up coil \cite{Chupp_1994}, SQUID magnetometer \cite{Gemmel2010} or a co-located alkali-metal magnetometer \cite{Walker2013,Walker2019}. Using alkali-metal vapor for in-situ polarization and readout of the nuclear spins allows for a compact sensor, but presents more challenges in avoiding interactions between spin species.
Our previous implementation of a \textsuperscript{3}He--\textsuperscript{129}Xe comagnetometer with \textsuperscript{87}Rb detection and decoupling achieved a rotational sensitivity of 7 nHz after 8 hours of averaging~\cite{Limes_2018}. However, it was limited by frequency shifts due to \textsuperscript{87}Rb decoupling pulses and the finite dynamic range of the Rb magnetometer.

In this Letter, we present a general experimental scheme, implemented using \textsuperscript{3}He and \textsuperscript{21}Ne spins, that enables direct measurements of the nuclear spin precession frequencies and overcomes both limitations. We implement a \textit{Ramsey scheme} in which nuclear spins precess entirely in the dark, free from optical or alkali-induced perturbations.  A  microwave field resonant with alkali-metal hyperfine transition suppresses the back-action of alkali polarization on nuclear spin precession. The Rb magnetometer is used to detect only a small residual spin precession signal due to Ramsey pulse frequency errors. This allows one to increase the nuclear spin polarization to extend the sensitivity of nuclear spin comagnetometers beyond the sensitivity of the Rb readout sensor.

The \textsuperscript{3}He--\textsuperscript{21}Ne comagnetometer is implemented using an 8-mm-diameter spherical cell made of GE180 glass, filled with 120\,Torr of \textsuperscript{3}He, 1.2\,atm of \textsuperscript{21}Ne, 40\,Torr of N\textsubscript{2}, and a droplet of enriched \textsuperscript{87}Rb located in the stem of the cell. The position of the Rb droplet can be adjusted via  thermal gradient applied between the cell and the stem to minimize long-range dipolar interactions between Ne and He arising from geometric asymmetries \cite{Limes_2019}.

Nuclear spin signals are detected using a pulse train \textsuperscript{87}Rb magnetometer~\cite{Limes_2018}. Short Rb $\pi$ pulses ($\pi_y$, $\pi_y$, $\pi_{-y}$, $\pi_{-y}$) are applied in sync with pump light polarization modulation ($\sigma^+$, $\sigma^-$, $\sigma^+$, $\sigma^-$). The Faraday rotation signal from a transverse probe laser is demodulated with a lock-in amplifier and gives signal proportional to $B_y$ field. At an operating temperature of 134\,$^\circ$C, the magnetometer achieves a sensitivity of 10\,fT$/\sqrt{\mathrm{Hz}}$, slightly higher than the thermal noise from magnetic shields~\cite{Lee2008}. The Rb $\pi$ pulse train causes frequency shifts for He and Ne spins with a quadratic dependence on the pulse amplitude. To eliminate these systematic effects, the Rb field $\pi$ pulses  and laser beams are adiabatically turned off during nuclear spin precession.

An oscillating Ramsey field resonant with each of the nuclear Larmor frequencies is applied along $\hat{y}$ and smoothly ramped up and down over $t_p = 40$~s to tip both spins by $\pi/2$ into the transverse plane, satisfying $\gamma_i B_i t_p = \pi$. The nuclear spins then precess freely in the dark for $\tau = 180$ s, as illustrated in Fig.~1. Subsequently, a second Ramsey tipping field is applied to generate an additional $\pi/2$ rotation, realigning the two nuclear spins near the $z$ axis in the direction opposite to their initial orientation. The tipping pulses, with frequencies of approximately 0.8 and 8 Hz, are generated by a Topping D90SE audio DAC card with an upgraded voltage reference and synchronized through a Singxer SU-2 audio bridge to an external rubidium clock.

\begin{figure}[b]
\centering
\includegraphics[width=0.45\textwidth]{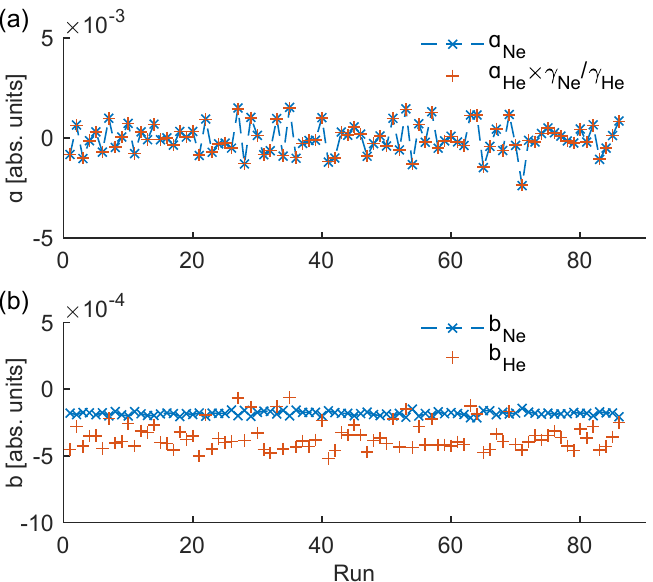}  
\caption{(a) The amplitude \(a\) of the sine component of \(A/A_0\) tracks field-induced frequency variations \(\delta\omega\) over time. (b) The amplitude \(b\) of the cosine component, arising from inaccuracies of  the tipping-field amplitude and detection noise. The offset originates from a small  error in the  tipping field.
}
\label{fig:NetipAMP}
\end{figure}

After the second Ramsey pulse, the pulse-train Rb magnetometer is re-engaged and the resulting small transverse spin components are detected: 
\begin{equation}
A(t) =\sum_i  A_0^i 
a_i\sin\!\left(\omega_{\mathrm{rf}}^i t+{\phi_\mathrm{d}^i}\right) +A_0^ib_i\cos\!\left(\omega_{\mathrm{rf}}^i t+\phi_\mathrm{d}^i\right)
\label{eq:freqamp}
\end{equation}
where $A_0^i$ is the full amplitude of nuclear spin $i$ during free precession, $\omega_{\mathrm{rf}}^i$ is the Larmor frequency of nuclear spins under the Rb pulse train~\cite{Limes_2018}. The phase $\phi_d$ is calculated so that a frequency detuning $\delta\omega_{i}$ causes a signal in the cosine component and a tipping amplitude mismatch $\delta B_i $ causes a signal in the sine component. Under these conditions, one can show that
$a_i=\left(\tau+4 t_{p}/\pi\right)\delta\omega_i$ and $b_i=\pi\delta B_i/B_i$ for small deviations~\cite{Wang_PhD}. During the experiment the fits are performed in real time and a  correction is calculated to adjust the frequency of the Ramsey pulses for the next shot so the amplitudes $a_i$ and $b_i$ remain small. Fig.~\ref{fig:NetipAMP} shows an example of the residual signal amplitudes in a typical run. One can see that $a_i$ amplitudes accurately represent drifts in the magnetic field and are highly correlated for He and Ne, while $b_i$ amplitudes are much smaller and their deviation is due to fitting noise and small errors in the setting of the tipping pulse amplitudes. 

To accurately extract He and Ne precession frequencies, we also correct for Block-Siegert shifts due to counter-rotating tipping field components and cross-talk between He and Ne tipping fields. These corrections can be calculated analytically and checked by direct integration of Bloch equations for He and Ne spins \cite{Wang_PhD}.

 At the end of the detection interval, we use active spin damping to flip both spins by 180$^{\circ}$. The signal of the Rb magnetometer is applied to $B_x$ coil, creating a $\pi/2$  phase shift for both spins. With appropriate feedback sign it flips both spins to the original orientation along $\hat{z}$ axis and then we smoothly turn off the gain of the damping feedback. To measure the  spin amplitude $A_0^i$ we apply a calibration pulse to tip the spins by  $2^\circ$. The size of the calibration angle is chosen to maintain linear response of the Rb magnetometer while generating a  large calibration signal compared to typical residual signals. After the calibration pulse, we switch the pumping laser to higher power and continuous $\sigma^{+}$ polarization to re-polarize nuclear spins. Before the next measurement cycle, we apply another spin damping sequence to initialize the spins precisely parallel to the $\hat{z}$ axis.

 \begin{figure}[b]
\centering
\includegraphics[width=0.4\textwidth]{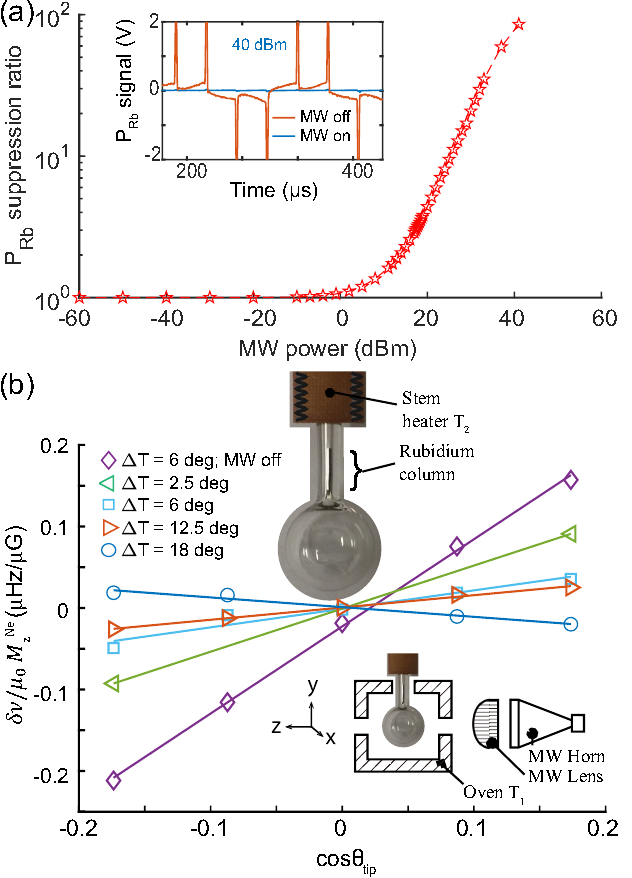}  
\caption{(a) Microwave-induced Rb depolarization ratio. In the experiment the MW power set to 40 dBm, suppressing back-action Rb polarization by a factor of 100. Inset: Pulse-train Rb magnetometer signals with and without the microwave field. (b) By scanning the angle of the first tipping pulse to leave a fraction of the nuclear spins along the $z$ axis, a frequency shift proportional to the longitudinal magnetization $M_z^{\mathrm{Ne}}$ is observed, arising from spin-spin  interactions.
Insets show the spherical cell, connected via a stem to a smaller upper bulb. The stem is sealed by a column of liquid rubidium. By precisely controlling the temperature difference $\Delta T=T_1-T_2$, the height of the rubidium column can be adjusted to finely tune $d_z$ of the cell.}

\label{fig:stemT}
\end{figure}

We identified several sources of systematic shifts in spin precession frequencies and developed methods to eliminate them. 
The first originates from Rb back-polarization $P^{\mathrm{Rb}}_{z}$ generated via spin exchange with He and Ne spins \cite{BenAmarBaranga1998,Chann_2002,Ghosh_2010},
 
\begin{align}
B_z^i &= \tfrac{2}{3} \kappa_{i}  \mu_B \mu_0 [\mathrm{Rb}] P^{\mathrm{Rb}}_z, \notag\\
P^{\mathrm{Rb}}_z &= 
\frac{3  \sigma^{se}_{\mathrm{NeRb}}\overline{v}_{\mathrm{NeRb}} }{\mu_\mathrm{Ne}\Gamma_{\mathrm{Rb}}}M^{\mathrm{Ne}}_z +\frac{ \sigma^{se}_{\mathrm{HeRb}}\overline{v}_{\mathrm{HeRb}}  }{\mu_\mathrm{He}\Gamma_{\mathrm{Rb}}}M^{\mathrm{He}}_z,
\label{eq:PRb}
\end{align}

where $\kappa_{i}$ are the Rb-noble gas contact interaction enhancement factors, $\kappa_{\mathrm{HeRb}}$ and $\kappa_{\mathrm{NeRb}}$ \cite{Ghosh_2010,Romalis_1998}, $\left[\mathrm{Rb}\right]$ is the density of Rb vapor, $\Gamma_{\mathrm{Rb}}$ is the longitudinal relaxation rate of Rb in the dark. To suppress this effect, a microwave field at \(6.835\,\mathrm{GHz}\), resonant with the ground-state hyperfine splitting of \({}^{87}\mathrm{Rb}\), is applied with  linear polarization along the \(\hat{x}\) axis throughout the dark evolution, effectively depolarizing residual rubidium polarization. Fig.~3a shows the degree of longitudinal Rb spin suppression achieved with the microwave field as a function of power. Using the microwave field to depolarize Rb dramatically reduces frequency shifts for nuclear spins.

Additional frequency shifts in the $^{21}$Ne and $^{3}$He precession originate from magnetic scalar and dipolar interactions between nuclear spins~\cite{VLASSENBROEK_1996,Limes_2019}. The equivalent field affecting nuclear spin precession is given by~\cite{Wetering2021}
\begin{equation}
\frac{1}{\mu_0}  \mathbf{B}^{\mathrm{He}}
= -\frac{3}{2}d_z M^{\mathrm{He}}_z \, \hat{\mathbf{z}} 
+ \left( -d_z + \frac{2}{3} \kappa_{\mathrm{HeNe}} \right) M^{\mathrm{Ne}}_z \, \hat{\mathbf{z}}.
\label{eq:fNeshift}
\end{equation}
and similar equation for Ne with He and Ne labels exchanged. The first term corresponds to long-range dipolar interactions of co-rotating components of mononuclear dipole field. It depends on the cell geometrical shape parametrized by the magnetometric demagnetizing factors, $\langle H_j\rangle_{V}=-n_j M_j$, (j = x, y, z). For a perfect sphere $n_j=1/3$. We introduce a cell asymmetry factor  $d_z=n_z-1/3$. Assuming cylindrical symmetry around the stem,  $n_x=1/3+d_z, n_y=1/3-2d_z$. The second term describes interactions between heteronuclear spins. In addition to long-range dipolar fields, it includes second-order scalar contact coupling $\kappa_{\mathrm{HeNe}}$ \cite{Vaara2019}, estimated to be on the order of $-1\times10^{-3}$ for $^3$He-$^{21}$Ne \cite{VaaraPC}.

We define the field-corrected co-magnetometer frequency $\nu_\mathrm{CM}=\nu_\mathrm{Ne}-(\gamma_{\mathrm{Ne}}/\gamma_{\mathrm{He}})\nu_\mathrm{He} $. The correction to this frequency from dipolar and scalar shifts is given by

\begin{equation}
\delta \nu_\mathrm{CM}=
 \frac{\mu_0\gamma_{\mathrm{Ne}}}{2\pi}
   \left(\frac{1}{2}d_z + \frac{2}{3} \kappa_{\mathrm{HeNe}} \right) 
   \left(M^{\mathrm{He}}_z-M^{\mathrm{Ne}}_z\right).
\label{eq:fHeshift}
\end{equation}

One can see that these systematic effects are also proportional to the longitudinal nuclear magnetization. Our experimental sequence minimizes average longitudinal magnetization during spin precession, since it is in the transverse plane during the dark interval and the longitudinal components approximately cancel between the two $\pi/2$ pulses. However,  changes in the magnetization due to relaxation or changes in the Rb density due to temperature drift will reduce the degree of cancellation. One can also adjust the relative duration of the first and second $\pi/2$ pulses to fine-tune the degree of cancellation.

To study these effects, we intentionally modify the tipping pulses so the magnetization has a finite $\hat{z}$ component during dark precession. We also shorten the pumping interval to cause nuclear magnetization to decay from one run to the next, particularly for Ne spins. In this way we can measure a correlation between the co-magnetometer frequency and Ne magnetization. The results of these measurements are summarized in Fig. 3b. After turning on the microwave field to eliminate Rb back-polarization effect, we vary the temperature difference between the cell and the stem. Reducing the stem temperature moves the molten Rb droplet and increases the volume occupied by polarized gas in the stem, which increases $d_z$ and compensates for a negative $\kappa_{\mathrm{HeNe}}$. 

Another possible source of systematic shifts is the quadrupolar energy splitting for $^{21}$Ne. However, unlike earlier experiments \cite{Chupp_1990}, we do not observe any evidence of quadrupolar splitting in $^{21}$Ne spin precession. One possibility is that we operate in the regime where the quadrupolar splitting is smaller than the quadrupolar relaxation rate. One can show \cite{WalkerPC, Wang_PhD} that under these conditions the resonance frequencies become mathematically degenerate.  After subtracting Earth's rotation rate, $\nu_\mathrm{CM}$ is on the order of $10^{-7}$~Hz, limited by the uncertainty in the  orientation of the bias field relative to Earth's axis.

\begin{figure}[thbp]
\centering
\includegraphics[width=0.45\textwidth]{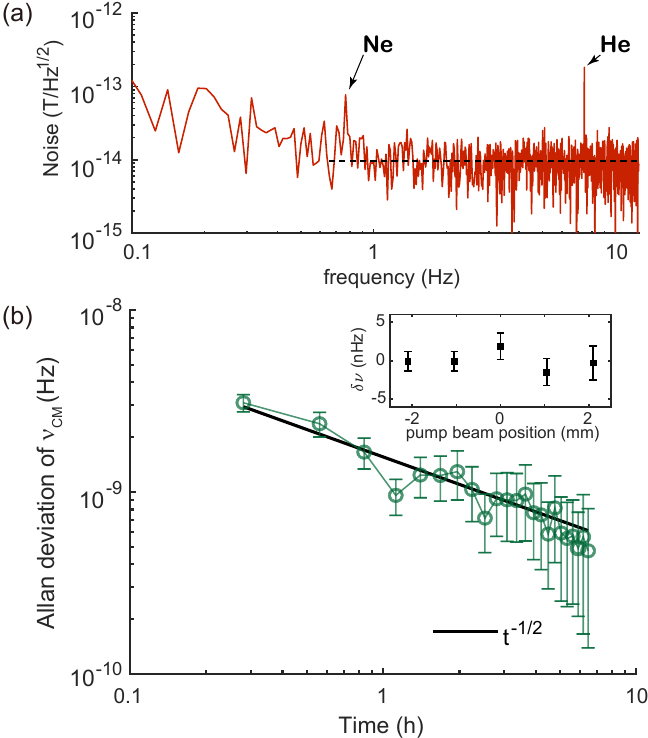}  
\caption{(a) Frequency spectrum of the pulse train Rb magnetometer, showing examples of small residual signals from Ne and He spin precession.  (b) Allan deviation of the co-magnetometer frequency, showing an upper stability limit of $580~\mathrm{pHz}$. The sensitivity to inertial rotation is equal to  $0.002^\circ/\sqrt{\mathrm{h}}$. Inset: Stability of the comagnetometer under deliberate pump laser misalignment from the cell center.}
\label{fig:allan}
\end{figure}

Fig.~\ref{fig:allan} shows the noise spectrum of the Rb pulse train magentometer and the Allen deviation for the co-magnetometer frequency $\nu_\mathrm{CM}$. The  comagnetometer achieves frequency sensitivity of $1.4~\mathrm{nHz}/\sqrt{\mathrm{h}}$ and a resolution of $580\ \mathrm{pHz}$ after 6~h integration without discernible drift,  representing more than an order-of-magnitude improvement over the best previously reported results with in-situ detection \cite{Limes_2018}. The inertial sensing performance with angle random walk (ARW) of $0.002^\circ/\sqrt{\mathrm{h}}$ and bias drift $<0.001^\circ/\mathrm{h}$ makes this device competitive with navigational grade ring laser gyros \cite{Savage2013}. To demonstrate the robustnesss of clock-comparison co-magnetometer, we vary the position of the pump laser beam across the cell by several mm, as llustrated on the inset in Fig.~\ref{fig:allan}b, and find no significant changes at 1~nHz level. 

The fundamental sensitivity of clock-comparison co-magnetometers is determined by the Cramér-Rao bound \cite{Gemmel2010}. If the co-magnetometer sensitivity is expressed in units of magnetic field, it can be related to the sensitivity of the signal detection magnetometer $\delta B_d$ \cite{Terrano_2022},
\begin{equation}
\delta B_{CM}=\frac{ 3\sqrt{6} }{ \gamma_\mathrm{Ne} \mu_0\kappa_{\mathrm{NeRb}} M_\mathrm{Ne} T}\delta B_d.
\end{equation}
The dimensionless gain factor $ \gamma_\mathrm{Ne} \mu_0\kappa_{\mathrm{NeRb}} M_\mathrm{Ne} T$ is proportional to the nuclear magnetization and the precession measurement time $T$ and is approximately equal to 100 for our conditions. While the experiment is currently not limited by the readout noise, the gain factor could be increased by using longer spin precession times and larger nuclear magnetizations. We find the transverse relaxation times  are equal to 8100\,s for \textsuperscript{21}Ne and 7200\,s for \textsuperscript{3}He, limited by quadrupole relaxation and magnetic field gradients, respectively, so much longer measurement times  are possible.

In summary, a Ramsey-type dual-species nuclear spin comagnetometer has been developed in which long-lived $^3$He and $^{21}$Ne spins precess in the dark and are subsequently detected by an in-situ Rb magnetometer.
Microwave field resonant with Rb hyperfine transition is used to eliminate Rb back-polarization with a minimum effect on nuclear spin frequencies.   By controlling the position of the Rb droplet in the stem of the cell we cancel the dipolar and scalar nuclear spin–spin interactions. Long-term stability and robustness of this comagnetometry technique is  promising for detection of spin-dependent interactions beyond the Standard Model without requiring fast signal modulation as well as for development of compact nuclear spin inertial rotation sensors. This work was supported by NSF award PHY-1912364.

\bibliographystyle{apsrev4-2}
\bibliography{references}

\end{document}